# Modelling radiation damage to ESA's Gaia satellite CCDs

George Seabroke*[a], Andrew Holland[a], Mark Cropper[b]
[a]Planetary & Space Sciences Research Institute, The Open University, Milton Keynes, UK;
[b]Mullard Space Science Laboratory, University College London, UK

## ABSTRACT

The Gaia satellite is a high-precision astrometry, photometry and spectroscopic ESA cornerstone mission, currently scheduled for launch in late 2011. Its primary science drivers are the composition, formation and evolution of the Galaxy. Gaia will achieve its scientific requirements with detailed calibration and correction for radiation damage. Microscopic models of Gaia's CCDs are being developed to simulate the charge trapping effect of radiation damage, which causes charge transfer inefficiency. The key to calculating the probability of a photoelectron being captured by a trap is the 3D electron density within each CCD pixel. However, this has not been physically modelled for Gaia CCD pixels. In this paper, the first of a series, we motivate the need for such specialised 3D device modelling and outline how its future results will fit into Gaia's overall radiation calibration strategy.

## 1. INTRODUCTION

Gaia will perform a survey of the entire sky, mapping all sources complete to $V$ = 20 mag[1]. From L2, the satellite will observe quasars, other galaxies, extra-solar planets, asteroids, comets and test general relativity and cosmology, but its primary science driver is the origin of our Galaxy. This will be achieved by observing $10^9$ stars throughout the Galaxy (1% of the Galactic stellar population). By measuring the positions of these stars tens of times over the 5-year baseline between the start of 2012 and the end of 2016, Gaia will derive their mean positions, parallaxes and proper motions. At $V$ = 15 mag, Gaia aims to measure these with accuracies of 25 µas, 25 µas and 25 µas yr$^{-1}$ respectively. Position and parallax (distance) define the 3D Galactic position of each star but proper motion only defines the 2D space velocity of each star in the plane of the sky. To fully unravel our Galaxy's formation history requires 3D space velocities: the missing component of the 6D phase space is radial velocity. This will be measured for the brighter stars by an onboard spectrograph with 1-15 km s$^{-1}$ accuracy.

Having introduced Gaia's science goals, we relate these goals to Gaia's focal plane and CCDs in Section 2. Section 3 introduces space radiation damage, how it affects the Gaia CCD pixel and the scientific consequences. Gaia's radiation calibration strategy is outlined in Section 4. We compare how existing models simulate 3D electron density within a pixel in Section 5 and conclude in Section 6 how these can be improved.

## 2. GAIA SATELLITE

### 2.1 Gaia focal plane

Gaia has two rectangular telescopes (1.45 × 0.5 m each), with viewing directions separated by 106.5°, which will feed the largest yet shared focal plane of CCDs (see Fig. 1). In order to survey the whole sky, the satellite will spin continuously and precess. As different images from the two telescopes transit the same focal plane, their undispersed light illuminates the Astrometric Field (AF) CCDs. This allows the angular separation of objects from each telescope to be continuously measured. As Gaia orbits the Sun each year, the simultaneous observations in the two viewing directions allow absolute rather than relative parallax to be determined. The final astrometric accuracies (see Section 1) will only be reached when the whole dataset is processed to yield the Astrometric Global Iterative Solution.

*g.m.seabroke@open.ac.uk

After transiting the AF CCDs, light from the images then passes through the low-resolution Blue Photometer (BP) and Red Photometer (RP) prisms, which disperse blue and red light in the same direction as the transit (along scan) onto the BP and RP CCDs respectively. As the transit continues, light from the images passes through the medium-resolution, integral-field Radial Velocity Spectrograph (RVS) grating, which disperses calcium triplet spectra along scan onto the RVS CCDs (see Fig. 1 and Table 1 for details).

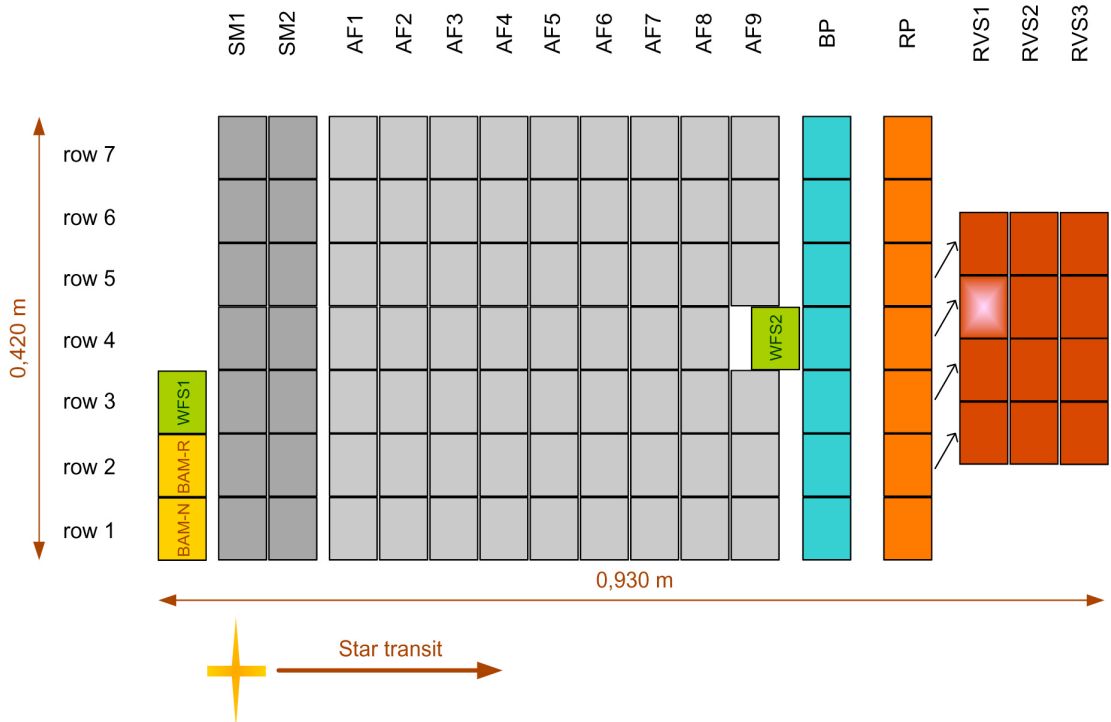

Fig. 1. The Gaia focal plane includes CCDs for Basic Angle Monitoring (BAM) between Gaia's two telescopes, Wave Front Sensing (WFS), two Sky Mapper (SM) strips to record the field of view from each telescope, Astrometric Field (AF), Blue Photometer (BP), Red Photometer (RP) and Radial Velocity Spectrograph (RVS) CCDs. (Courtesy of F. Safa, EADS-Astrium.)

Table 1. Summary of the Gaia focal plane instruments and science goals.

| Instrument | λ (nm) | Dispersion (nm/pixel) | Measurement | Astrophysical (stellar) parameters |
|---|---|---|---|---|
| AF | 330-1050 | Not dispersed | Astrometry | Positions, parallaxes, proper motions |
|  |  |  | Photometry | Apparent magnitudes |
| BP | 330-680 | 3-30 | Spectro-photometry | Effective temperatures, gravities, metallicities, line-of-sight extinction to individual stars, alpha particle enhancement in some stars |
| RP | 640-1050 |  |  |  |
| RVS | 847-874 | 0.02698 | Spectroscopy | Radial velocities. $V \leq 14$ mag stars: emission lines and abundances as well as more accurate radial velocities |

## 2.2 Gaia CCD pixel

The Gaia CCDs (CCD91) were designed and are currently being manufactured by e2v technologies (UK). The pixel architecture is one of the most complex ever built. In order to minimise the number required to fill the Gaia focal plane, the Gaia CCDs are large area devices (~26.5 cm$^2$). All the CCDs are back-illuminated: the AF and BP CCDs are 16 μm deep, while the red-enhanced RP and RVS CCDs are 40 μm deep with a special coating. The pixel size samples the AF PSF in the AL (along-scan) direction (10 μm), while maximising the full well capacity in the AC (across-scan) direction (30 μm). As the satellite will spin continuously, the pixels will be clocked in Time Delay and Integration (TDI) mode at the same speed as images scan along the focal plane, allowing images to integrate as they transit each CCD (4.42 s). There are 4500 AL pixels (and 1966 AC pixels) so the integration time for each pixel is ~1 ms. This integration time is further sub-divided to match the different sizes of the four electrodes in each pixel (see Fig. 2). Firstly, voltage is applied to electrodes ϕ1 and ϕ2 for 0.3 ms. During this time, signal-generated photoelectrons are attracted to accumulate in a charge packet in the normal buried channel (BC) and supplementary buried channel (SBC) regions beneath these electrodes. Then voltage is stopped to ϕ1 but applied to ϕ2 and ϕ3 for 0.2 ms. This moves the charge packet from underneath ϕ1 and ϕ2 to underneath ϕ2 and ϕ3 and the integration proceeds for 0.2 ms. Then voltage is stopped to ϕ2 but applied to ϕ3 and ϕ4 for 0.3 ms and so on. This keeps the closest possible match between the motion of the integrating measured image and the incoming optical image.

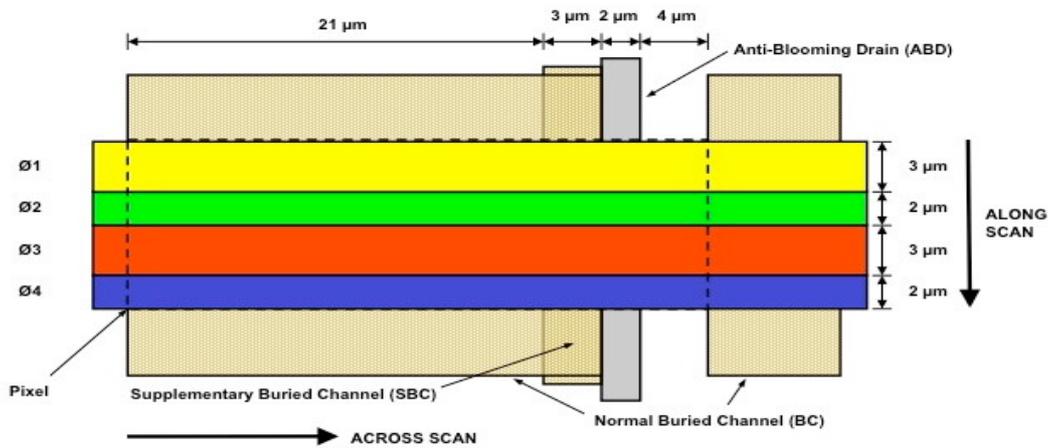

Fig. 2. Front-side schematic of a Gaia pixel showing the four electrodes (ϕ).

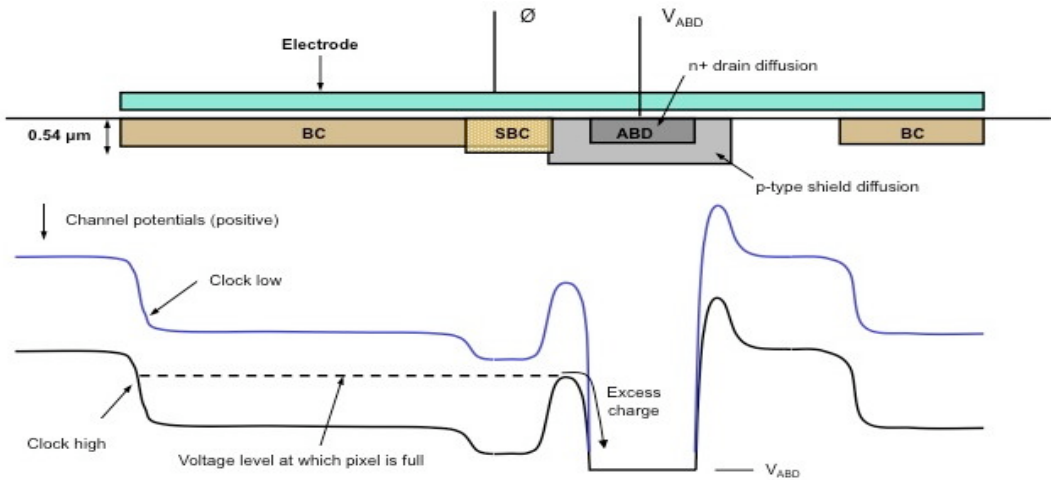

Fig. 3. *Top:* Schematic of the vertical cross-section through Fig. 2 in the across-scan direction. *Bottom:* Channel potential profile in the same direction, resulting from different voltages being applied to the electrode (clock low = 0 V, clock high = 10 V).

# 3. RADIATION DAMAGE

## 3.1 Trap formation

Space radiation damage has been identified as a mission-critical issue for Gaia. Gaia will operate from L2, which is an interplanetary radiation environment. This includes high-energy extra-solar cosmic rays but the damaging radiation particle fluence is dominated by lower energy solar protons from solar flares at solar maximum, which coincides with Gaia's launch. Due to the size of Gaia's focal plane and the limited launch mass budget of the Soyuz Fregat, only minimal radiation shielding is incorporated in Gaia's design. Therefore, most of these particles hit the Gaia CCDs. Non-ionising, displacement damage occurs when sufficiently energetic protons knock silicon atoms out of their lattice positions across the entire CCD (see Fig. 2). The resulting vacancies (V) can move around the lattice until they combine with other atoms, e.g. phosphorous (P) or oxygen (O), or combine with other vacancies in the BC and SBC to form electron traps, with different discrete energy levels between the valence and conduction bands (see Table 2). High thermo-mechanical stability is required in order for Gaia to achieve its specified astrometric accuracy and so in situ heat treating (annealing[2]) the traps to break their bonds to remove them from the CCDs is not an option. Gaia's operating temperature of 163 K was chosen as a compromise to minimise the overall effect of the different trap species (see Table 2 and Equations 1 and 2).

Table 2. Summary of approximate trap properties[3]. $n_t$ is the trap density after a radiation fluence of 5×10$^9$ 10 MeV protons cm$^{-2}$ (approximately Gaia's expected end-of-life fluence). $N_t$ is the resulting number of traps active in the AL direction under two electrodes in a Gaia imaging pixel within the charge packet storage volume = (21+3) × 5 × 0.54 µm (see Figs 2 and 3. A charge packet will not be able to encounter all of these traps because fringing fields and column isolation (see Fig. 2) mean the volume that the electrons sit in is less than the BC+SBC volume underneath two electrodes. The other symbols are defined in the text accompanying Equations 1 and 2.

| Trap name | Bond(s) | $E$ (eV) | $\sigma X$ (cm$^2$) | $n_t$ (cm$^{-3}$) | $N_t$ | $\tau_r$ (ms) | Active area on CCD |
|---|---|---|---|---|---|---|---|
| Si-E centre | P-V | 0.46 | 6×10$^{-15}$ | 1.00×10$^{11}$ | 6.8400 | 6×10$^5$ | Imaging area (AL) |
| Divacancy | V-V | 0.41 | 5×10$^{-16}$ | 1.55×10$^{10}$ | 1.0044 | 2×10$^5$ | Imaging area (AL) |
| Unknown | ? | ~0.3 | 5×10$^{-16}$ | 2.00×10$^{10}$ | 1.2960 | 8×10$^1$ | Imaging area (AL) |
| Multiple V | V-V-V | 0.21 | 5×10$^{-16}$ | 1.55×10$^{10}$ | 1.0044 | 1×10$^{-1}$ | Readout register (AC) |
| Si-A centre | O-V | 0.17 | 1×10$^{-14}$ | 2.00×10$^{11}$ | 12.960 | 4×10$^{-4}$ | Readout register (AC) |

## 3.1 Trap physics

If a charge packet encounters a trap, the trap can capture an electron from the charge packet and release it at a later time. These processes are stochastic and are described by the following equations derived from Shockley-Read-Hall theory[4,5]. The probability ($P_c$) that an empty trap will capture an electron in time $t$ is given by

$$P_c = 1 - e^{\frac{-t}{\tau_c}}, \text{ where } \tau_c = \frac{1}{\sigma v n}, \text{ where } v = \sqrt{\frac{3kT}{m}}, \tag{1}$$

where $\tau_c$ is the exponential capture time constant, $\sigma$ is the trap capture cross-section, $v$ is the electron thermal velocity, $n$ is the electron density in the vicinity of the trap, $k$ is the Boltzmann constant, $T$ is the temperature and $m$ is the effective electron mass (in silicon: ~0.5 free electron rest mass). The probability ($P_r$) that an occupied trap will release its electron in time $t$ is given by

$$P_r = 1 - e^{\frac{-t}{\tau_r}}, \text{ where } \tau_r = \frac{e^{\frac{E}{kT}}}{X \chi \sigma v N}, \text{ where } N = 2\left(\frac{2 \pi m k T}{h^2}\right)^{\frac{3}{2}}, \tag{2}$$

where $\tau_r$ is the exponential release time constant, $E$ is the trap energy level below the conduction band, $X$ and $\chi$ are entropy and field enhancement factors respectively (both ~1), $N$ is the effective density of states in the conduction band and $h$ is the Planck constant.

The effect of traps on Gaia observations is determined by how quickly traps are filled by a passing charge packet and how similar their subsequent $\tau_r$ values are to the TDI period (the charge packet dwell period under two electrodes: 200-300 ms – see Section 2.2). We address the former issue later in the paper by analysing Equation 1. Table 2 shows the results of evaluating Equation 2 to calculate $\tau_r$ for each type of trap. These are given in units of ms to compare to the TDI period. Clocking of pixels in the across-scan direction in the CCD readout (RO) register is $\sim 1 \times 10^{-4}$ ms (RO period). The V-V-V and O-V traps have $\tau_r$ values that are both less than the TDI period. Consequently, if these traps capture electrons from passing charge packets within the TDI period, they will also release these electrons within the TDI period so charge packets will be left unaltered in TDI mode. However, the $\tau_r$ values for these traps are both greater than the RO period so the charge packet will be altered in RO mode. The P-V, V-V and the 0.3 eV traps all have $\tau_r$ values greater than the TDI period and so charge packets will be altered in TDI mode by the presence of these traps. Therefore from Table 2, the total number of active traps under two electrodes in the BC and SBC of a Gaia imaging pixel after a radiation fluence of $5 \times 10^9$ 10 MeV protons cm$^{-2}$ (approximately Gaia's expected end-of-life fluence) is ~9 or ~18 per pixel (4 electrodes) or ~81 000 within the 4500 pixels along one scan of the CCD or $\sim 2 \times 10^8$ within the $\sim 9 \times 10^6$ pixels in the imaging section of a Gaia CCD (excluding RO pixels).

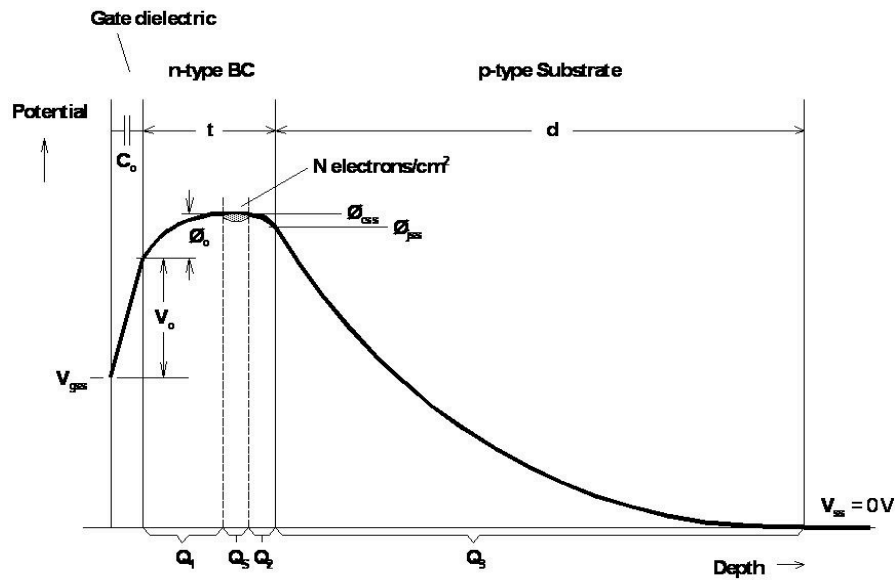

Fig. 4. Schematic of a CCD pixel's vertical channel potential profile through the BC (t = 0.54 μm) and substrate (d). The depletion region (t+d) in AF/BP and RP/RVS CCDs is 12 and 35 μm respectively.

However, charge packets will not meet all of these traps because their electrons do not fill the entire BC and SBC volumes. Instead, they sit in the channel potential maxima. Fig. 3 shows that when the signal is less than the SBC capacity (1300 electrons for Gaia), these electrons sit in the SBC potential maximum, which is a smaller volume than the BC potential profile. Therefore, these small signals being transferred along scan in the SBC meet a smaller number of traps than larger signals being transferred along scan in the BC, which is the reason for the SBC's inclusion in the pixel architecture. Charge packet volume is approximately independent of signal size, until the signal is sufficiently large to alter the shape of the potential distribution. Fig. 4 illustrates a small charge packet sitting in the channel potential maximum. Simulations[6] show that within the potential, electrons in a small charge packet are confined to move over a depth of 0.15 μm. This is much less than the BC depth of 0.54 μm, as illustrated in Fig. 4. Of course, not all the traps a charge packet actually encounters will alter it because a trap can only hold one electron. Thus, if a trap is full as the charge packet passes, the packet will not be altered. Only if the trap is empty is it able to capture an electron and alter the passing charge packet.

The easiest way to visualise this alteration to charge packets is to consider how the P-V and V-V traps affect Gaia photometric AF, BP, RP and RVS measurements. These are 'slow' traps, meaning their $\tau_r$ values are much longer than the TDI period. Because all traps are stationary but the charge packets are moving in TDI mode, electrons trapped by the P-V and V-V traps will not be emitted as trailing charge until all the charge packets making up an image have completely passed by. Hence, electrons can be completely removed from the point spread function (PSF) of a passing image (charge loss), increasing the charge transfer inefficiency (CTI). This is the usual measure of the effect of radiation damage on an image. The 0.3 eV trap causes a more complex example of damage. Because its $\tau_r$ value is more similar to the TDI period (see Table 2), electrons are removed from charge packets at the front of the PSF and released into packets at the back of the PSF as they clock past these traps. It is this asymmetric charge redistribution and the corresponding centroid shift (or bias) that is the most relevant measure of radiation damage for Gaia astrometry and its precise image centroiding requirements[7]. As the BP/RP and RVS spectra are dispersed along 60 and 1104 pixels respectively, bias and charge loss due to the 0.3 eV trap will cause these spectra to change shape. Early analysis of Astrium test data from irradiated RVS CCDs show that charge loss is more important than bias[8]. This is because charge loss reduces the equivalent widths of stellar absorption lines, removing radial velocity information, which increases the random errors of the derived radial velocities.

## 4. GAIA RADIATION CALIBRATION STRATEGY

The Gaia radiation calibration strategy[9] follows five lines of attack:

1. Minimising radiation damage by further CCD hardware optimization (e.g. charge injection or diffuse optical background strategies to keep the P-V and V-V traps filled);
2. Astrium CCD hardware testing;
3. Analysis of Astrium CCD hardware tests by Astrium and members of the Gaia Data Processing and Analysis Consortium (DPAC);
4. Implementation of radiation damage effects in simulation software to produce simulated damaged data to compare with hardware results;
5. Software calibration in the astrometric, photometric and RVS data reduction pipelines.

These five lines of attack are mutually dependent and either feed into or require different models of radiation damage. Because of the vast amount of data that will need to be processed, the actual treatment of raw Gaia data will need to be very computationally efficient, i.e. using simplified, parameterised models of how images are affected by radiation damage. The development of these models is following the philosophy of starting as complex and as accurate as possible in microscopic models before simplifying (to reduce CPU time), while maintaining the required accuracy in macroscopic models.

Microscopic models are based on known physical processes and CCD properties, minimising arbitrary *ad hoc* assumptions. These model the charge transfer at sub-pixel level and the individual traps capturing and releasing electrons according to Equations 1 and 2. These models can be used to perform Monte Carlo simulations of the TDI and RO processes and should be able to reproduce hardware test results.

Macroscopic models represent the mean or statistical behaviour of a microscopic model. For example, rather than modelling the exact position of individual traps like in a microscopic model, a macroscopic model may be sufficiently accurate after modelling the effective number of traps along a CCD column. The current macroscopic model[10] in the Gaia data processing pipeline is analytical and represents a phenomenological description of various effects observed in hardware test data, rather than being a true physical model like a microscopic model. It is a 'forward modelling' approach, where the observed photoelectron counts are compared with modelled counts calculated by means of a suitable Charge Distortion Model (CDM). It treats the effects of radiation closest to the source of damage: correcting the biases introduced by the distortions of the PSF/LSF[11]. Radiation damage corrected image data is never calculated – only corrections to derived parameters are calculated. The CDM parameters are adjusted iteratively until the best fit is obtained between the observed and model counts, resulting in an estimation of the model parameters including those of the astronomical object like position and flux.

The input to the CDM is a sequence of values ($s_i$) representing a sampled image in successive pixels ($i$). The output is a sequence of distorted counts ($d_i$), representing the charge image. The two sequences are related by

$$d_i = s_i - c_i + r_i, \tag{3}$$

where $c_i$ and $r_i$ are the number of electrons captured and released in the current pixel respectively. The main part of the CDM is specifying how $c_i$ and $r_i$ depend on the current and previous pixel values. This could in principle be a function of a large number of variables. These could be simulated in multiple Monte Carlo runs of a microscopic model. However, the macroscopic model has to run quickly and so has to be simplified. This is done by assuming $c_i$ and $r_i$ are only functions of a small number of state variables that are computed recursively. The current first-order model uses just a single state variable called the equivalent fill level. This is the minimum level of $s_i$ required for part of the charge packet to be captured by traps.

By analysing the output charge distribution from a known input charge distribution (e.g. charge injection or a bright star), the average number of available traps and $\tau_r$ can be parameterised[12,13]. $\tau_r$ can be empirically determined by fitting exponentials to measured trap release curves. These parameters are inputs to the CDM charge capture and release models. The CDM charge capture model includes parameters representing $\tau_c$ that require external calibration. This needs to be done using a microscopic model that can reproduce all the hardware test results. Equation 1 shows that $\tau_c$ critically depends on electron density in the vicinity of a trap ($n$). The next section shows how this variable remains the most arbitrary ad-hoc assumption used in the various microscopic models developed for Gaia.

## 5. 3D ELECTRON DENSITY MODELS

In order to apply Equations 1 and 2 to a CCD simulation, it is necessary to model the volume and density of charge packets. There are two different types of models that can do this: confinement volume (or volume-driven) models and density distribution (or density-driven) models[14]. Confinement volume models assume charge packets have a finite confinement volume that increases as more electrons are added. They also assume that electron density is generally high enough for trapping within the charge packet volume to be considered instantaneous. Therefore, according to confinement models, all empty traps within the charge packet will capture an electron whilst those outside cannot and thus the volume of the charge packet drives the amount of trapping. These models are able to reproduce observed trapping as a function of signal size.

Analysis of CCD hardware tests[15] revealed the impact of diffuse optical background (DOB) on trap occupancy has implications for confinement volume models. The tests found that increasing the DOB from 0.3 to 5 electrons/pixel reduced the measured charge loss from ~30 to ~10% at $G$ = 18 mag, suggesting a very low level of DOB can have a significant and apparently disproportionate effect on trap occupancy. A further increase in DOB from 5 to 10 electrons/pixel had only a very small additional effect, suggesting the effect of DOB saturates by keeping one or two trap species full while leaving others unaffected. The effect of DOB on measured centroid shift was less significant than charge loss, suggesting that the DOB electrons could keep traps with the longest $\tau_r$ (slow traps) filled, whilst traps with shorter $\tau_r$ (fast traps) remain empty.

According to confinement volume models, a 5-electron charge packet (due to DOB) must have an extremely small volume and so cannot encounter enough traps in order to fill them and explain the hardware test results. In order to fill the numbers of traps observed in test data, a few DOB electrons would have to occupy a volume many thousands of times larger than an equivalent number of signal electrons. To explain how a small number of electrons appears to be able to fill an entire population of traps, it was considered that the electrons are not confined to a volume which increases with the number of electrons. Instead, density distribution models assume that the electrons are confined within the same geometrical volume of silicon regardless of their number and that only the electron density increases as more electrons are added. Therefore, all charge packets encounter the same number of traps but $P_c$ depends on the electron density in the vicinity of the trap, which increases with signal size.

The electron density of a DOB signal of 5 electrons distributed over the entire confinement volume will be very low. Consequently, $P_c$ will be extremely small during one TDI period and the DOB signal is unlikely to fill any traps during this short timescale. However, DOB is present all the time and so the probability that a trap captures a DOB electron on longer timescales is much higher. According to this model, any trap that is located such that the local electron density due to DOB gives rise to a $\tau_c$ that is shorter than its $\tau_r$ tends to remain full and so subsequent charge packets can pass by these traps unaltered. Thus the density of the charge packet drives the amount of trapping and provides a more plausible mechanism to explain the effects of DOB than the volume-driven models, while still being to reproduce observed trapping as a function of signal size.

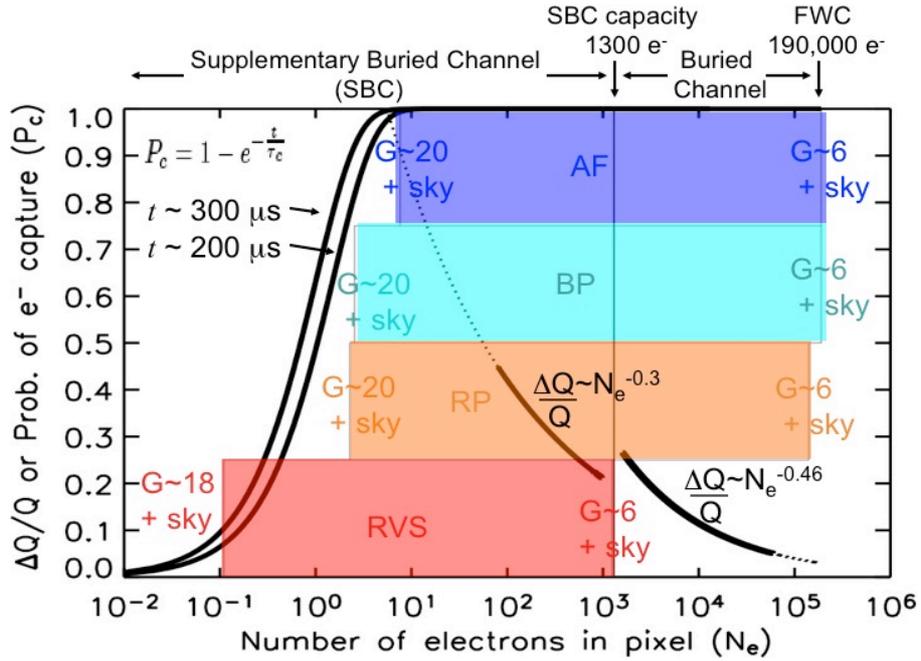

Fig. 5. *Left curves*: Probability of electron capture in a single TDI period as a function of the number of electrons in a charge packet in a pixel. *Right curves*: Power-law fit to fractional charge loss ($\Delta Q/Q$) after 18 000 TDI transfers with a 400 TDI delay between charge injection and signal on an e2v CCD97-72 device, measured by Hopkinson et al.[16], scaled to Gaia's average end of life radiation dose ($4.8 \times 10^9$ 10 MeV protons cm$^{-2}$). *Coloured blocks*: Range of each Gaia instrument relating the maximum and minimum $G$ magnitude to the resulting number of electrons in a pixel at the end of the CCD (after 4.42 s integration time including typical sky background surface brightness – see text for details of calculation). When there is 0.1 electron in a pixel, this means there will be 1 electron in 1 out of every 10 pixels.

Fig. 5 illustrates the density-driven probability of electron capture for the Gaia instruments. We have evaluated Equation 1 for the 0.3 eV trap for the two different TDI periods by assuming the value of the geometrical volume of silicon within which charge packets can reside. Volume becomes dependent on signal size when the signal is sufficiently large to alter the shape of the potential distribution. The volume before and after the onset of this behaviour can only be derived accurately by solving for the potential (see Fig. 4), using Poisson's equation, and the charge density, using the charge continuity equation, simultaneously. Nevertheless, we approximate the SBC electron confinement volume as 5 μm along scan (see Fig. 2) minus 1 μm (to account for fringing fields), times 3 μm across scan (see Fig. 3), times 0.15 μm in depth (see Fig. 4). The electron density is then simply approximated by the number of electrons divided by our assumed volume. Fig. 5 shows that $P_c \rightarrow 1$ within the SBC capacity so the BC volume does not need to be approximated.

The coloured blocks in Fig. 5 delineate the Gaia instrument bright and faint limits in terms of individual charge packet size in each pixel. The photometry of each instrument is defined in its own band. These can be most easily compared using the relation: $G_{AF} = G_{BP} = G_{RP} = G_{RVS} = V = 0$ mag for an unreddened Vega-like star (A0V). Using the magnitude zero-points for each band, which includes the telescope transmission (mirror reflectivity, mirror rugosity and mirror contamination), CCD quantum efficiency and transmission of the instrument optics, the total integrated flux for the bright and faint limits (including typical sky background surface brightness) were calculated. The number of pixels that will be assigned to each image to sample it fully (a Gaia 'window') was used to calculate the number of electrons in each pixel if the PSF was uniform. This assumption overestimates the number of electrons in the wings of the PSF and vastly underestimates the number of electrons in the PSF core.

Hopkinson et al.[16] injected an irradiated CCD with different sized charge packets to measure the fractional charge loss (ΔQ/Q). They measured ΔQ/Q using the first pixel response method – measuring the difference in signal between the first line to be readout and subsequent lines (for which a steady equilibrium had been established). We have plotted their power-law fits to their measurements as solid lines in Fig. 5. Their data is from CCD columns close to the output amplifier to avoid probing trapping due to serial transfer across the readout register. We have extrapolated these curves to extend over the Gaia range (dotted lines). This extrapolation suggests there will be complete loss of signal at the faint ends of the Gaia instruments, but there are many reasons why this will not happen. Firstly, all ΔQ/Q plotted in Fig. 5 occurs only in the first line. Subsequent lines have a smaller ΔQ/Q because the traps have already been filled by charge in the first line. Secondly, $P_c < 1$ for small charge packets so large charge packets will have more electrons captured than small charge packets. Thirdly, regular charge injection can keep slow traps filled, further reducing ΔQ/Q.

Fig. 5 shows that the SBC reduces ΔQ/Q at signals below ~1300 electrons compared to the BC. The average trend of the experimental data smoothly connects the two solid power-law curves, albeit with a relatively large scatter (not plotted in Fig. 5). This complicated behaviour over a small range of charge packet size is supported by 2D modelling of the SBC in the e2v CCD43 device, which shows that signal spills out of the SBC into the BC in complex geometries[17].

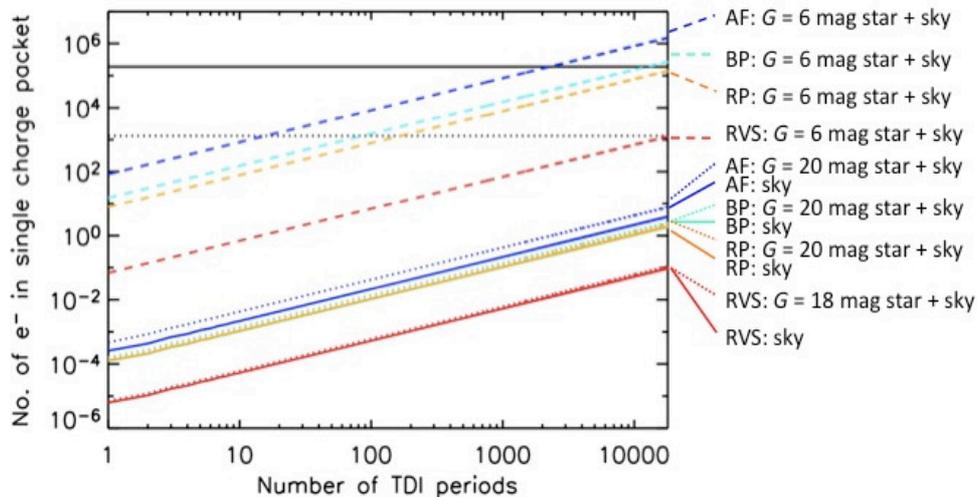

Fig. 6. Number of electrons in a single charge packet as a function of the number of TDI periods for each Gaia instrument: AF (blue), BP (cyan), RP (orange), RVS (red). The lines denote the following as seen by each instrument: typical sky background surface brightness (solid); faintest star windowed, $G = 20$ mag for AF, BP, RP, $G = 18$ mag for RVS (dotted); brightest star windowed, $G = 6$ mag (dashed). The black lines represent the full well capacities of the BC (solid) and SBC (dotted). The small steps in the lines at small numbers of TDI periods is due to the alternate TDI periods (300 and 200 ms).

The simple analytical 3D electron density model, illustrated in Fig. 5, suggests that $P_c \to 1$ when a charge packet consists of ~10 or more electrons. Fig. 5 shows $P_c \neq 1$ only occurs at the BP, RP and RVS faint ends. However, this relation to the Gaia instruments only applies instantaneously to charge packets in the last pixel at the end of an entire transit of one CCD. Microscopic models simulate the integrated alterations at the end of each transit by evaluating Equations 1 and 2 at each TDI period. Fig. 6 shows how the charge packet linearly increases its number of electrons in TDI mode along a CCD. Figs 5 and 6 highlight that at the AF, BP and RP bright ends, charge packets are always in the $P_c \rightarrow 1$ regime (>10 electrons) and are in the BC (and SBC) rather than only confined to the SBC. In contrast, the RVS bright end has ~100 TDI periods in the $P_c \neq 1$ regime before $P_c \rightarrow 1$. Only charge packets in the PSF core of spectra at the RVS bright end will have enough electrons to spill out of the SBC into the BC. Therefore, the vast majority of RVS charge packets will traverse each CCD entirely confined to the SBC.

Most microscopic models developed for Gaia have concentrated on estimating how radiation damage will affect astrometric accuracy. The first such model by L. Lindegren[7] simulated electron density within an old Gaia pixel design that did not include a SBC to estimate the astrometric bias and charge loss. Hardy et al.[18] used a specialist 2D device simulation package to construct a 3D model of potential and electron density for a Tektronix TK512 CCD 27 μm square pixel, by performing simulations of two 2D cross-sections at right angles through the device. Because these distributions are approximately Gaussian when the density is small, Lindegren analytically modelled electron density by multiplying the signal size by a normalised 3D Gaussian density function. The standard deviations in each dimension are approximated according to the architecture of the old Gaia pixel design. Lindegren developed an ad-hoc analytical model for when the charge density approached the BC doping concentration. The total charge was allowed to grow by increasing the width of the density distribution in all three dimensions.

Like the previous model, the next electron density model for Gaia by M. Robbins[3] was not derived for the current Gaia pixel design but for an e2v CCD42-10 pixel. A bespoke internally-developed device simulation software was used to model the 2D potentials and electron densities. In static imaging mode, the simulation of ΔQ/Q agreed well with hardware test data. However, the simulation seemed to overestimate ΔQ/Q in TDI mode by a factor of 2 to 3. Possible reasons proposed by M. Robbins for these differences are the limitations and uncertainties of using 2D distributions to derive 3D charge capture and/or Equations 1 and 2 do not describe charge capture and release at very small signals.

The Brunel University microscopic TDI model[19] was developed to simulate the astrometric accuracy of bright AF stars. Figs 5 and 6 show that their assumption of instantaneous electron capture is valid for these stars. This is because at the AF bright end, charge packets always consist of >10 electrons. Fainter AF stars that begin the CCD transit with <10 electrons accumulate >10 electrons in a small number of TDI periods compared to the total transit. This has the advantage that Equation 1 did not need to be evaluated and the 3D electron density distribution did not need to be assumed or derived.

The first microscopic TDI simulation to model the final Gaia pixel architecture was by A. Short[20,14]. Like the Lindegren model[7], rather than calculating the theoretical 3D electron density distribution according to physical principles, Short models the electron distribution in each pixel dimension with arbitrary, flexible functions for each direction that can be adjusted to give the best fit to the hardware test data. The electron density is then given by the number of electrons multiplied by the electron probability distributions in each dimension. His model assumed a Gaussian electron probability distribution in CCD depth. Previous CCD simulations[6] and Fig. 4 suggest this is approximately correct for all BC CCDs. The standard deviation of this vertical Gaussian, and a parameter that describes the flattening of the potential well as more electrons are added, are both free parameters fitted to the hardware test data. The model assumes that the along and across scan electron distributions are uniform within their confinement lengths. This assumption approximates the BC potential and electron distributions in Fig. 3 to a top-hat distribution extending over <21+3 μm (effective pixel width across scan). The Short model is the only one to model the SBC. For small signals, the electron distribution within the SBC is assumed to be Gaussian, where its standard deviation is a free parameter fitted to the hardware test data. At large signals, the SBC potential is assumed to collapse to give a simple, uniform electron distribution across the effective pixel width across scan. An exponential is used to model the transition between confinement in the SBC at small signals and BC confinement within the effective pixel width across scan at large signals. Short's Monte Carlo model, with >10 free parameters is able to partly reproduce hardware test data of how radiation alters a PSF. However, it is computationally very slow and so it is not practical to incorporate fitting algorithms into it to find the parameters that give the best fit to the data.

# 6. CONCLUSIONS AND FUTURE WORK

Gaia's preliminary design review was completed in June 2005. Around half the Gaia CCDs have been built by e2v. This paper briefly introduces Gaia's primary science goals but Gaia's expected scientific harvest is of almost inconceivable extent and implication. This harvest is based on unprecedented positional and radial velocity measurements. However, detailed calibration and correction for radiation damage is required. The current radiation calibration strategy for Gaia is to correct this damage with simplified, parameterised macroscopic models. Parameters in these models, especially those relating to charge capture, need to be calibrated by more detailed models that simulate the microscopic physics of charge trapping and release. In this paper, we have reviewed all the microscopic models developed for Gaia. We conclude that there remain arbitrary, *ad hoc* assumptions that are not sufficiently physically motivated. The primary example of this is that previous microscopic models have either assumed a 3D electron density distribution, modelled it in 2D for a non-Gaia pixel or fitted a parameterised 3D electron density distribution to hardware tests. Three dimensional electron density distributions as a function of the number of electrons in a charge packet have not been specifically modelled for the Gaia CCD pixel architecture. Nevertheless, it is possible to physically model this parameter in detail using specialised 3D device modelling software.

This type of modelling is required to reduce uncertainties in the microscopic TDI models currently being developed for Gaia[21,22] by reducing the free parameter space of Equations 1 and 2. It is critical for two reasons. Firstly, to model the volume of charge packets as a function of their number of electrons to give the number of traps, for a given trap density, that the charge packet will encounter as it moves through the CCD. This is important for simulations throughout the Gaia instrument range in Figs 5 and 6 when the charge packets are confined only within the SBC and when they are large enough to fill both a part of the BC and the whole of the SBC. Secondly, modelling the density of charge packets as a function of their number of electrons holds the key to accurately calculating the probability of a photoelectron being captured by a trap. Although this is not required for the PSF core of bright AF stars, the PSF core of faint stars in all the Gaia instruments and parts of all PSF wings will have <10 electrons according to Fig. 6 and so in these cases the probability of a photoelectron being captured by a trap needs to be realistically modelled. Also, if traps are empty when a charge packet of any size transits the CCD, most damage will occur in the first pixel when the number of electrons will often be <10. Figs 5 and 6 show that the density is most important for the faint end of the AF, BP and RP instruments and critical for a large range of the RVS instrument, when the charge packets are confined only within the SBC. An open question for the RVS instrument is whether it can successfully observe down to $G = 18$ mag. Will the very small number of electrons in the charge packets of these stars survive a CCD transit to be readout above the noise?

The microscopic models being developed should be able to reproduce CCD hardware tests at very low signal levels. However, there may be insurmountable technical reasons why the current test set up cannot conduct CCD experiments that can reach the low signal levels expected in the RVS instrument. In this case, these microscopic models may be the only information available on the behaviour of the RVS instrument at its faint end.

Therefore, we are currently modelling the Gaia CCD pixel architectures using the semi-conductor industry simulation software standard for Technology Computer Aided Design (TCAD), called Silvaco. This requires defining the doping concentration in the BC and SBC and then solving for the channel potential (Poisson's equation) and the charge density (charge continuity equation) simultaneously. By inserting different sized charge packets into the simulation, we will be able to derive the 3D volume and electron density as a function of the number of electrons in the packet. Analytical functions will be fit to these distributions so they can be used to solve Equation 1 in microscopic Gaia TDI simulations. The accuracy with which these simulations will be able to reproduce the hardware test data and calibrate macroscopic models (that will correct for the radiation damage) may ultimately be limited by any variation between pixel properties within a single CCD and between different CCDs that could result from e2v production spreads. This may be the reason for the scatter around the empirical fits (not shown in Fig. 5) measured by Hopkinson et al.[16] in their fig. 11. However, ultimate limits to modelling could also come from small number trap statistics and/or statistical variations in proton exposure from column to column in the CCD. Future hardware testing should clarify these issues.

## ACKNOWLEDGEMENTS

This paper has benefited enormously from very useful discussions with David Burt and Mark Robbins (e2v), David Katz (Gaia CU6 Spectroscopic Processing) and with the participants of the first Gaia Radiation Task Force Meeting in Cambridge, especially Alex Short (ESTEC), Thibaut Prod'homme (ELSA student at Leiden) and all the members of the Gaia CU5 (Photometric Processing) DU10 (PSF and LSF Calibration) group: Floor van Leeuwen, Duncan Fyfe, Michael Davidson, Nigel Hambly, Nicholas Cross and Patricio F. Ortiz.